\documentstyle[11pt,emulateapj,epsfig,psfig]{article} 
\lefthead{Turolla \& Dullemond}
\righthead{ADVECTION--DOMINATED FLOWS}

\begin{document}

\title{Advection--dominated Inflow/Outflows from Evaporating Accretion Disks}

\author{R. Turolla}
\affil{Department of Physics, University of Padova, \\
Via Marzolo 8, I--35131 Padova, Italy \\
e--mail: turolla@pd.infn.it}
\and
\author{C.P. Dullemond}
\affil{Max Planck Institut f\"ur Astrophysik,\\
Karl Schwarzschild Strasse 1, D--85748 Garching, Germany\\
e--mail: dullemon@mpa-garching.mpg.de}

\begin{abstract}

In this {\em Letter} we investigate the properties of
advection--dominated accretion flows (ADAFs) fed by the
evaporation of a Shakura--Sunyaev accretion disk (SSD). In our
picture the ADAF fills the central cavity evacuated by the SSD and
extends beyond the transition radius into a coronal region. We
find that, because of global angular momentum conservation, a
significant fraction of the hot gas flows away from the black hole
forming a transsonic wind, unless the injection rate depends only
weakly on radius (if $r^2\dot\sigma\propto r^{-\xi}$, $\xi< 1/2$).
The Bernoulli number of the inflowing gas is negative if the
transition radius is $\lesssim 100$ Schwarzschild radii, so matter
falling into the hole is gravitationally bound. The ratio of
inflowing to outflowing mass is $\approx 1/2$, so in these
solutions the accretion rate is of the same order as in standard
ADAFs and much larger than in advection--dominated inflow/outflow
models (ADIOS). The possible relevance of evaporation--fed
solutions to accretion flows in black hole X--ray binaries is
briefly discussed.
\end{abstract}


\section{Introduction}\label{sec-intro}

In recent years X--ray and optical observations provided
increasing evidence that advection--dominated flows (ADAFs,
Ichimaru \cite{ichimaru:1977}; Narayan \& Yi
\cite{narayanyi:1995b}; Abramowicz et al. \cite{abrchen:1995}) are
a ubiquitous feature of accretion at all scales, from black holes
(and possibly neutron stars) in X--ray binaries (e.g.~Narayan
\cite{narayan:1996}; Esin, McClintock \& Narayan
\cite{esinmcclnar:1997}; Menou et al. \cite{menou:1999}) to
supermassive black holes in the center of galaxies (e.g. Narayan,
Yi, \& Mahadevan \cite{nayima95:1995}, Lasota et al.
\cite{lasota:1996}, Narayan et al. \cite{nar98:1998}; Gammie,
Narayan, \& Blandford \cite{ganarbla:1999}).

Despite their structure being inherently two--dimensional, much
theoretical work on ADAFs still relies on a vertically--integrated
approach, which, although questionable, may indeed catch some of
the essential properties of the model, as its successful
application to various sources shows. The standard ADAF picture
was, however, challenged in a recent paper by Blandford, \&
Begelman (\cite{blabe:1999}). Since, at least in self--similar
ADAFs, the Bernoulli number is positive (as already noted by
Narayan, \& Yi \cite{narayanyi94:1994}, \cite{narayanyi:1995a}),
they suggested that a large outflow may form. If this is the case,
they have shown that the mass carried out by the wind must exceed
by orders of magnitude that which is crossing the horizon, turning
ADAFs into Advection--Dominated Inflow/Outflow Solutions, or
ADIOS. Although the positiveness of the Bernoulli number is only a
necessary (but not sufficient) condition for starting an outflow,
this argument poses a serious problem  to the ADAF model for black
hole accretion and must be addressed carefully.

Goal of this {\em Letter} is to investigate how, and to which
extent, the inclusion of the source of ADAF material affects the
Bernoulli number and the onset of a wind. Advection--dominated
flows in black hole X--ray binaries (BHXBs) are most likely
produced by the evaporation of a Shakura--Sunyaev disk (SSD,
Shakura, \& Sunyaev \cite{shaksuny:1973}), as observational and
theoretical arguments suggest (e.g.~Narayan, Mahadevan, \&
Quataert \cite{namaqua:1999}). The evaporation process is not
fully understood as yet (see e.g.  Meyer, \& Meyer--Hofmeister
\cite{meyermeyhof:1994}; Honma \cite{honma:1996}; Dullemond
\cite{dul:1999}; R\'oza\'nska \cite{roz:1999}) and definitely
needs to be modeled in at least two spatial dimensions. Here we
just assume a simple, analytical law for the evaporation rate and
limit ourselves to a vertically--integrated description of the hot
flow. A quite general argument, based on angular momentum
conservation, indicates that purely inflowing solutions can not
exist if the accretion rate decreases with radius. Numerical
models, computed for several values of the $\alpha$--viscosity
parameter and of the transition radius $R_0$, support this
conclusion. In all of them a stagnation radius (where the radial
velocity vanishes) separates an inner inflowing region from an
outer transsonic wind. We find that the Bernoulli number for the
infalling gas is negative if the transition radius is less than
$\sim 100$ Schwarzschild radii. In these solutions about $1/3$ of
the mass carried inwards from large radii by the thin disk reaches
the horizon. More extended inflows, with $R_0 \gtrsim 100 R_s$,
have a region of  positive Bernoulli number, and are likely to be
replaced by an ADIOS.

\section{The Model}\label{sec-model}

We consider a highly idealized model for a hybrid accretion flow,
in which a hot, advection--dominated phase coexists with a
Shakura--Sunyaev disk. All the ADAF gas is assumed to be supplied
by the evaporation  of the surface layers of the SSD which extends
down to the transition radius $R_0$. In the following we will not
be concerned with the SSD anymore and focus our attention on the
hot component. The inner rim is chosen to be at $R_{in}= 3GM/c^2$
($M$ is the hole mass), and $R_0>R_{in}$. The advection--dominated
flow is described by the usual stationary, vertically--integrated
equations (see e.g.~Narayan, Kato, \& Honma
\cite{narkathon:1997}), which now include the energy and momentum
exchange between the hot gas and the evaporating material; the
standard $\alpha$--prescription for viscosity is retained, the
pseudo--Newtonian potential is used to describe the gravitational
field and we neglect radiative losses. The physics of the
evaporation process is still unclear, so we just assume that
matter is lost from the SSD (per unit area and time) according to
the simple law
\begin{equation}\label{sigmadot}
\dot\sigma(R) = \left\{\matrix{\dot\sigma_0\,(R/R_0)^{-\xi-2} & \quad &
             R\ge R_0\cr
              0 & \quad & R<R_0\, . } \right.
\end{equation}

The continuity equation reads
\begin{equation}\label{eq-contin}
\frac{d (R\Sigma v_R)}{dR} = R\dot\sigma
\end{equation}
which can be immediately integrated to yield $-2\pi R\Sigma v_R =
\dot M_{ADAF}(R)$.
In the previous expressions $\Sigma$ is the surface density and
$v_R$ the radial velocity, chosen to be negative for matter
flowing towards the hole. The ADAF accretion rate now depends on $R$
and is given by

\begin{equation}\label{mdot-adaf}
\dot M_{ADAF}(R) =
\left\{\matrix{\dot M\;(R/R_0)^{-\xi}
                      - \dot M_{out}  &\; R\ge R_0\cr
                      \dot M  - \dot M_{out}
                      &\; R<R_0 } \right.
\end{equation}
where $\dot M_{out}=(2\pi\ R\Sigma v_R)|_{R\to\infty}$, $\dot
M=2\pi\xi^{-1}\dot\sigma_0R_0^2=\dot M_{SSD}(R\to\infty)$ is total
accretion rate, and we consider
only the case $\xi>0$, so $\dot M_{ADAF}$ decreases with $R$.
If $\dot M_{out}>0$, $\dot M_{ADAF}$ becomes negative for $R>R_{st} =
[ 1+ (\dot M -\dot M_{out})/\dot M_{out}]^{1/\xi}R_0$ and the gas
crosses the horizon
at a rate $\dot M_{in}=\dot M -\dot M_{out}<\dot M$; this implies that
also $v_R$ has to switch sign at the stagnation radius $R_{st}$.

In the following we assume that the injected material rotates at the
Keplerian angular speed $\Omega_K$ and that the rising gas elements move
predominantly in the vertical direction with velocity $\ll \Omega_KR$. Since
the ADAF is expected to rotate at $\Omega<\Omega_K$,
the difference in the circular
speeds produces a torque. Conservation of angular momentum then implies

\begin{equation}\label{eq-angmom}
\Sigma R v_R \frac{d(R^2\Omega)}{dR}-\frac{d}{dR}\left(\Sigma\nu
R^3\frac{d\Omega}{dR}\right)=\dot\sigma (\Omega_K-\Omega)R^3
\end{equation}
where $\nu=2/3\alpha c_s H$ is the viscosity coefficient, $H$  the flow
half--thickness and $c_s$ the isothermal sound speed.
Other possible sources of friction between the SSD and the ADAF like, e.g.,
magnetic stresses, have been neglected.

The ADAF has to spend part of its energy to heat up the injected
gas which has initially the same temperature of the
Shakura--Sunyaev disk. At the
same time heat is produced by frictional dissipation. Including
both these effects, the local energy balance takes the form

\begin{eqnarray}\label{eq-energy}
&&v_R\left[(\gamma-1)\frac{dc_s^2}{dR} - c_s^2
 \frac{d\ln(\Sigma/H)}{dR}\right] -
\nu\left(R\frac{d\Omega}{dR}\right)^2 = \nonumber\\
&& \frac{\dot\sigma}{\Sigma}
\left[\frac{1}{2}v_R^2 + \frac{1}{2} R^2(\Omega_K-\Omega)^2
-\frac{\gamma}{\gamma -1}c_s^2\right]
\end{eqnarray}
where  $\gamma$ is the adiabatic index and we assumed that the
rising gas elements do not suffer any energy loss before thermalizing with the
hot plasma.

Finally, the radial force balance is expressed as

\begin{equation}\label{eq-radmom}
v_R\frac{dv_R}{dR}+(\Omega_K^2-\Omega^2)R + \frac{dc_s^2}{dR} +
c_s^2 \frac{d\ln(\Sigma/H)}{dR}= -\frac{\dot\sigma}{\Sigma}\,v_R\, ;
\end{equation}
the last term arises because the injected mass has zero radial
momentum and is usually negligible. As expected, Eqs.
(\ref{eq-contin}), (\ref{eq-angmom})--(\ref{eq-radmom}) reduce to
the standard ADAF form for $\dot\sigma=0$ .

\section{Global Solutions for the Hot Flow}\label{sec-global}

It can be easily shown that, under the usual assumptions, the flow
equations admit a self--similar solution. However, at variance
with standard advection--dominated models, a self--similar regime
is not always allowed for evaporation--fed ADAFs.
In fact, at least if $0<\Omega<\Omega_K$, angular momentum conservation
implies $(1-2\xi)/v_R < 0$, so inflowing self--similar solutions
are ruled out unless $\xi<1/2$. The reason for this is as follows.
If the amount of mass injected into the ADAF drops off too
quickly with increasing $R$, viscous stresses can not transport
enough angular momentum outwards because the density becomes too
low at large radii. The ADAF can not get rid of its angular
momentum by transferring it to the SSD either, since the cold disk
rotates faster. On a physical basis, it seems therefore unlikely
that purely inflowing global solutions could exist. The
only possibility for the flow to transport efficiently angular
momentum to infinity is to advect it, reversing its motion from a
certain radius onwards.

Eqs. (\ref{eq-contin}),
(\ref{eq-angmom})--(\ref{eq-radmom}) have been solved numerically
using a Henyey relaxation method (Dullemond, \& Turolla
\cite{dultur:1998}). Despite several attempts, no purely
inflowing solution was found for
$\xi > 1/2$, as the previous argument predicts.
In all models both the outflow and the inflow are
transsonic. The presence of three critical points (the two sonic radii and
the stagnation radius) reduces the number of the boundary condition from
five (there are three first order and one second order differential
equations) to two, the other three being replaced by regularity conditions
at the critical points. Since the viscosity must be well--behaved at both
the inner and outer edge (the no--torque condition, see Narayan, Kato, \& Honma
\cite{narkathon:1997}), we require
\begin{equation}\label{bcs}
\frac{d\log\Omega}{d\log R} = -2\qquad \ \hbox{at} \ R=R_{in} \
\hbox{and} \ R=R_{out}\, .
\end{equation}
This choice introduces no further degree of freedom.
The solution depends only on the function $\dot\sigma(R)$, on
$\alpha$ and $\gamma$. Both the mass loss rate at large radii $\dot
M_{out}$, and the stagnation radius follow from the calculation and are
found to obey the analytical expression for $R_{st}$ derived in
\S \ref{sec-model} to high accuracy.

We have computed several series of models with $0.1<\alpha< 1$,
$1/2<\xi<3/2$ and $\gamma=3/2$, varying $R_0$ in the range
$20R_s$--$1000R_s$. As Eqs. (\ref{eq-angmom})--(\ref{eq-radmom})
show, $v_R$, $\Omega$ and $c_s$ are independent of $\dot\sigma_0$
and $M$; the same is for the ratio $\dot M_{in}/\dot M_{out}$ of
accretion to mass loss rates. The radial dependence of $v_R$ and
$c_s$ for a typical run is plotted in Fig. \ref{plot-rad-var};
solutions with different values of $\alpha$ and $\xi$ show the
same general behaviour. Only a limited $\xi$--range was considered
here, but we have verified that steeper (e.g. exponential)
injection laws give quite similar results. The  accretion to mass
loss ratio goes from $\dot M_{in}/\dot M_{out}\sim 0.5$ to $\sim
0.3$ as $R_0$ increases from 20 to 500 $R_s$ and is not much
dependent of $\alpha$ and $\xi$. Correspondingly, $R_{st}$ is in
all cases about $1.5 R_0$. As expected, for $R_0\gtrsim 500R_s$
the inflow closely resembles standard ADAFs and, in particular,
has a nearly self--similar region at intermediate radii. For
smaller values of $R_0$ the accretion flow has no room to attain
self--similarity, being squeezed in between the sonic and the
stagnation radius.

\section{Discussion}\label{sec-discuss}

Our solutions share with ADIOS the property that a significant
(although very different) fraction of the material is expelled in
a wind. The two models, however, differ substantially in many
respects.

Blandford, \& Begelman (\cite{blabe:1999}) constructed
self--similar advection--dominated solutions for which the
Bernoulli number $Be$ is negative assuming that $\dot
M_{ADAF}\propto R^p$ with $p>0$. They concluded that all the mass
which does not reach the horizon escapes in a wind more massive
than the ADAF by orders of magnitude. Although no detailed model
including the wind has been presented as yet, preliminary
hydrodynamical calculations seem indeed to support the original
suggestion that the mass inflow rate increases with radius (Stone,
Pringle, \& Begelman \cite{stopribe:1999}). In addition, 2--D
simulations by Igumenshchev, \& Abramowicz (\cite{iguabra:1999})
have shown that ADIOS--like solutions are present for large
$\alpha$. In our model the outflow is a direct consequence of
having assumed that the ADAF material is supplied by the SSD. The
accreting gas has to transfer angular momentum to larger radii
and, since $\dot M_{ADAF}$ is decreasing with increasing $R$ in
the evaporation region, this can not be done by viscous stresses
alone. From the stagnation radius onwards, the gas is
centrifugally accelerated away from the hole, carrying angular
momentum with it: no stationary solution would be possible without
an outflowing region. This is not related to the positiveness of
the Bernoulli number. Moreover, the wind is not produced by the
ADAF itself but originates directly from the evaporating SSD
material.

The basic objection Blandford and Begelman raised to standard
ADAFs is very general and concerns also the inflowing part of
evaporation--fed models. The Bernoulli number for our solutions is
shown in Fig. \ref{plot-rad-be}. In the wind region the Bernoulli
number becomes positive shortly beyond the stagnation radius,
while $Be$ can be either positive or negative for the infalling
gas, depending on $R_0$. Solutions with a small transition radius
have always $Be < 0$. This can be understood considering the
energy the hot flow transfers to the cold injected material (last
term in eq. [\ref{eq-energy}]). The evaporation stops at $R_0$ but
the inflowing gas will stay cooler (with respect to a standard
ADAF) a bit further down. For $R_0\lesssim 100 R_s$ viscous
dissipation has no time to heat the flow sufficiently before it
crosses the horizon, keeping $Be$ negative 
(see also Nakamura \cite{naka:1998} for a discussion on the sign of 
the Bernoulli number
in non--self--similar ADAF solutions). As $R_0$ increases
above $\sim 100R_s$ the Bernoulli number is still negative close
to $R_0$ but it flips sign as soon as the inflowing gas has become
hot enough. An ADIOS--type wind may be expected where $Be>0$.

The inclusion of a source for the hot gas points towards the
existence of a more general class of advection--dominated flows,
with somewhat intermediate characteristics between standard ADAFs
and ADIOS. In the light of the relation to (and possible inclusion
of) ADIOS--type flows, we refer to these models as ``Consistent
Inflow And Outflows'', or CIAOs.

Our vertically--integrated CIAO model has to be supported by
detailed 2--D and 3--D hydrodynamical calculations, but the
present analysis suggests that for $R_0\lesssim 100 R_s$ the
inflowing matter is gravitationally bound and no ADIOS--like wind
is expected to be present. CIAOs can supply the central hole with
about 1/3 to 1/4 of the mass originally carried inwards by the thin disk.
This may be relevant for applications of advection--dominated
flows to BHXBs and AGNs. It has been often proposed (e.g. Narayan
\cite{narayan:1996}; Esin, McClintock, \& Narayan
\cite{esinmcclnar:1997}; Belloni, et al. \cite{belloni:1997};
Narayan, Mahadevan \& Quataert \cite{namaqua:1999}) that the
different spectral states observed in BHXBs may be explained in
terms of a bimodal disk in which the transition radius varies with
time. The inferred values of $R_0$ are in between few $R_s$
and $\approx 100\, R_s$. Recent optical/UV observations also
indicate that the inner edge of the thin disk in the nucleus of
M81 and NGC 4579 lies at $\sim 100\, R_s$ (Quataert, et al.
\cite{quaetal:1999}). Although a more detailed analysis is
definitely required before any firm conclusion can be drawn, it is
interesting to note that the maximum value of the transition
radius for which our models have $Be <0$ is close to the observed limit
mentioned above. CIAOs may comfortably provide the hot accretion
flow in all the range of $R_0$ implied by observations. At the same time,
the absence of hot flows with an extent $\gtrsim 100R_s$ can be interpreted in
terms of the onset of a strong wind which, as in ADIOS, blows off the
accreting part of the CIAO and reduces dramatically the accretion rate 
onto the black hole.

\newpage

\centerline{
\epsfxsize=9.truecm
\centerline{{\epsfbox{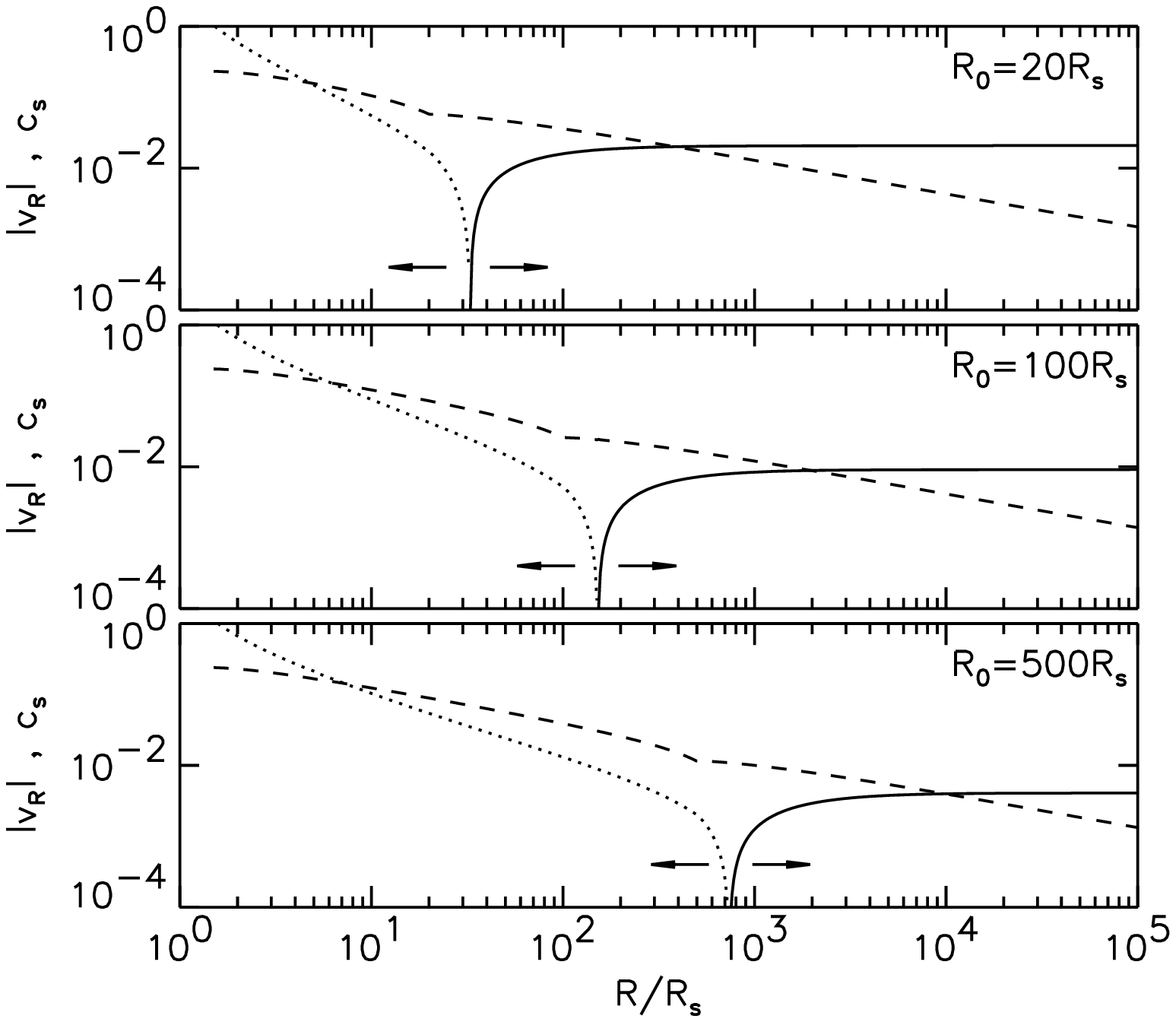}}}}

\figcaption[f1.ps]{The radial run of $|v_R|$ (dotted/full line for
negative/positive
values) and $c_s$ (dashed line), both in units of $c$,
for $\alpha=0.5$, $\xi=0.75$ and three values of $R_0$. From top to bottom
$\dot M_{in}/\dot M_{out} = 0.44,\, 0.37$ and 0.35. The arrows mark the
radial flow direction.
\label{plot-rad-var}}

\mbox{}
\vspace{1em}

\centerline{
\epsfxsize=9.truecm
\centerline{{\epsfbox{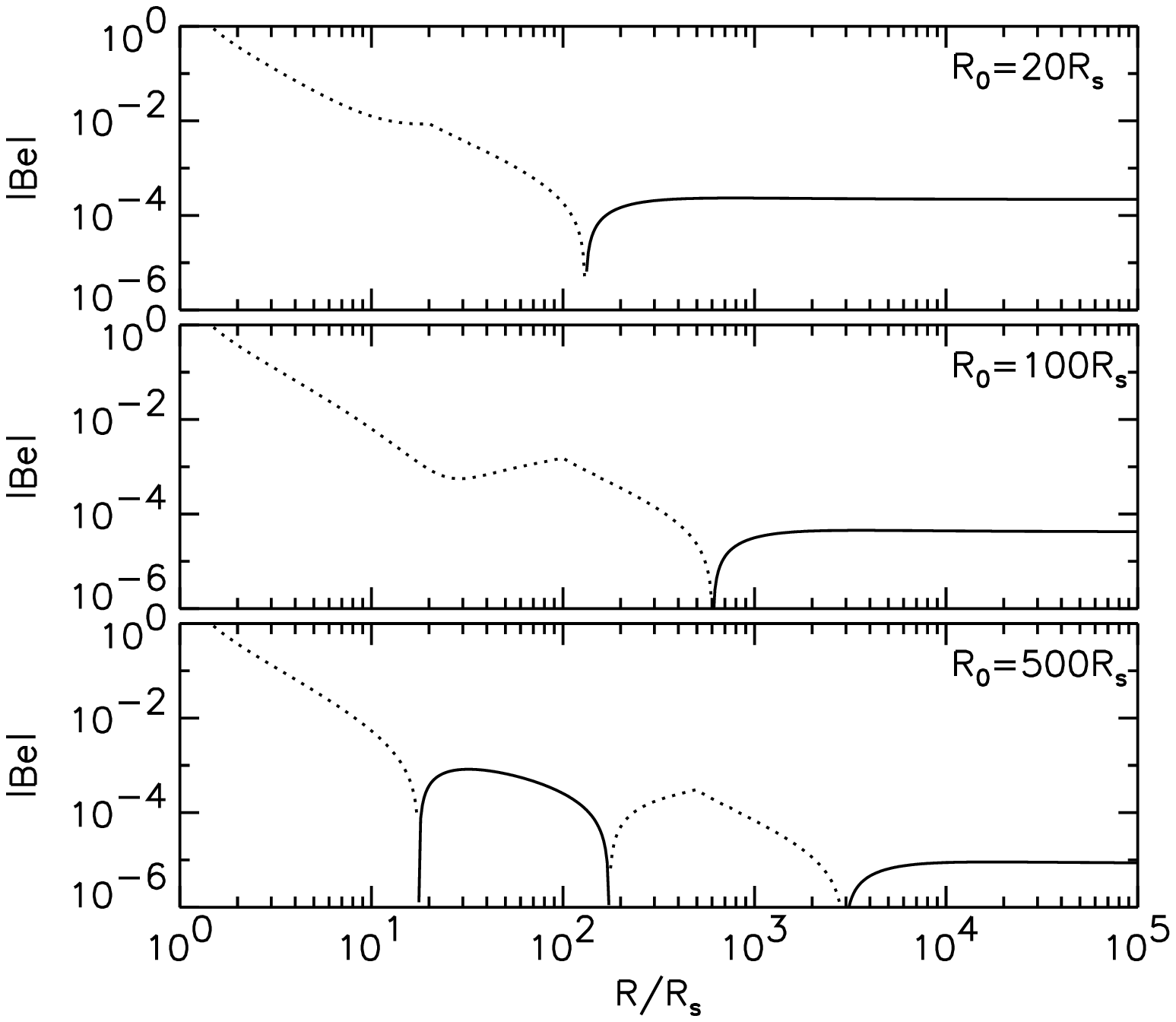}}}}

\figcaption[f2.ps]{Same as in fig. \ref{plot-rad-var} for $|Be|$ (in
units of
$c^2$). $Be$ is negative for $R<R_{st}$ in the middle panel, but
the dip at $\sim 20R_s$ signals that it is going soon to switch sign in
the inflowing region.
\label{plot-rad-be}}

\end{document}